\documentstyle[preprint,eqsecnum,aps]{revtex}
\tightenlines

\def\dfrac#1#2{{\displaystyle\frac{#1}{#2}}}
\begin{document}
\draft
\preprint{HEP/123-qed}
\title{
Carrier density change in Colossal Magnetoresistive Pyrochlore Tl$_{2}$Mn$_{2}$O$_{7}$
}
\author{H. Imai,\cite{byline} Y. Shimakawa, Yu. V. Sushko,\cite{byline2} and Y. Kubo}

\address{
Fundamental Research Laboratories, NEC Corporation, 34 Miyukigaoka,
 Tsukuba 305-8501, Japan
}

\date{\today}
\maketitle
\begin{abstract}
Hall resistivity and magneto-thermopower have been measured for colossal magnetoresistive Tl$_{2}$Mn$_{2}$O$_{7}$ over wide temperature and magnetic-field ranges. These measurements revealed that a small number of free electron-like carriers is responsible for the magneto-transport properties.  In contrast to perovskite CMR materials, the anomalous Hall coefficient is negligible even in the ferromagnetic state due to negligibly small skew scattering. The characteristic feature in Tl$_{2}$Mn$_{2}$O$_{7}$ is that {\it the carrier density changes with temperature and the magnetic field}. The carrier density increases around $T_{C}$ as the temperature is lowered or as the magnetic field is increased, which explains the CMR of this material. 
The conduction-band-edge shift, which is caused by the strong $s$-$d$ interaction between localized Mn moments and $s$-like conduction electrons, is a possible mechanism for the carrier density change.
\end{abstract}

\narrowtext

Colossal magnetoresistance (CMR) observed in carrier-doped perovskite manganese oxides, e.g. La$_{1-x} A_{x}$MnO$_{3}$  ($A$ represents a divalent elements such as Ca, Sr, Ba, or Pb), has created an active new field of physics, where the strong correlation between spin, charge, and lattice degrees of freedom plays an important role ~\cite{ref1,ref2,ref3,ref4}.
 The fundamental CMR behavior and the phase transition from a paramagnetic semiconductor to a ferromagnetic metal have been explained on the basis of a double-exchange model ~\cite{ref5}.
  The doped holes mediate the ferromagnetic interaction of adjacent spins, resulting in metallic conduction in the ferromagnetic state.
  To quantitatively describe CMR, though, the strong lattice coupling, i.e. the contribution of static and dynamical Jahn-Taller distortion of Mn$^{3+}$ ion must be taken into account ~\cite{ref6}. 
  
A pyrochlore manganese oxide, Tl$_{2}$Mn$_{2}$O$_{7}$, exhibits CMR behavior (MR ratio $-86~\%$ at 8~T) comparable to that of perovskite materials near $T_{C} \simeq 120$~K~\cite{ref7}.
   However, neither the double-exchange model nor the Jahn-Teller distortion assisted mechanism, are  applicable to describe the CMR in Tl$_{2}$Mn$_{2}$O$_{7}$.
 The carrier density is $\sim 0.005$ electrons per formula unit ~\cite{ref7,ref10}, which is too small to mediate the double exchange interaction between Mn spins.
 Jahn-Teller distortions are not expected due to inactive Mn$^{4+}$ ions~\cite{ref8,ref9}, and no structural anomaly was observed across $T_{C}$~\cite{ref11,ref12}. 
 The negative pressure dependence of $T_{C}$ also conflicts with the double-exchange model ~\cite{ref12a,ref13}. 
 In addition, the electronic band structure calculation for ferromagnetic Tl$_{2}$Mn$_{2}$O$_{7}$ revealed that a small number of carrier is produced in one of the down-spin bands hybridized among the Tl $6s$, Mn $3d$, and O $2p$ orbital~\cite{ref14a,ref14}.  
 Thus, we proposed an alternative scenario to explain this novel CMR effect~\cite{ref10,ref14}: The ferromagnetic ordering occurs by means of the superexchange interaction between Mn$^{4+}$  ions {\it via} O $2p$ and Tl $6s$ orbital, separately from the conduction process.
  This ferromagnetic ordering changes the electronic state drastically, and as a result, a spin polarized $s$-like carrier is produced below $T_{C}$. This provides the high conductivity in the ferromagnetic state.

In this Letter, we report the temperature dependence of the Hall resistivity and the magneto-thermopower. 
These properties exhibited characteristic behavior, in particular around $T_{C}$, and were strikingly different from those observed in a perovskite system. 
The results of the Hall resistivity and the thermopower measurements revealed that a small number of free-electron-like carriers govern the transport properties and that the carrier density abruptly increases around $T_{C}$ when the temperature is lowered or a magnetic field is applied. 

Polycrystalline samples were synthesized by solid-state reaction under high pressure. (Details were described in a previous paper~\cite{ref11}.) 
 Isothermal Hall resistivity measurements were carried out in magnetic fields up to 8~T by a four-probe method in the van der Pauw configuration at temperatures from 10 to 300~K. 
  Magneto-thermopower measurements in magnetic fields of 0, 2, 5, and 7~T, were done using a standard constant $\Delta T$ method in the temperature range of 4 to 300~K. 

In Fig. 1, we show Hall resistivity, $\rho_{xy}$, data as a function of the magnetic field at various temperatures. In a ferromagnet, $\rho_{xy}$ is written by the general formula,
\begin{equation}
\rho_{xy}=R_{O}B + 4 \pi M R_{S},
\end{equation}
where $R_{O}$ is the ordinary Hall coefficient, $B$ is the applied magnetic field, $R_{S }$ is the anomalous Hall coefficient, and $M$ is magnetization~\cite{ref15}.
In a free-electron model, the carrier density is obtained by $n=R_{O}/e$. 
 In the measured $T$ and $B$ ranges, the observed $\rho_{xy}$ is negative, indicating that the predominant conduction carrier is the electron. 
 The second term concerns the anomalous Hall effect, which is related to the skew scattering of itinerant carriers by a localized moment through spin-orbit interaction~\cite{ref15}. 
 We consider the contribution of $R_{S }$ to $\rho_{xy}$ in Tl$_{2}$Mn$_{2}$O$_{7}$ to be rather small, because the skew scattering is expected to be intrinsically weak. 
 Since the orbital angular momentum $(L)$ of Mn$^{4+}$ is quenched, the possible spin-orbit coupling exists between the  spin $(S)$ of the Mn $3d$ localized moment and $L$ of itinerant $s$-character carriers. 
 In this case, $L$ of carriers in the $s$-band is proportional to $k^{2}$  for small values of $k$ (i.e. small carrier numbers), giving rise to weak skew scattering~\cite{ref16}. 

The Hall resistivity data show three types of behavior in respect to temperature.
 In region A, where $T \le T_{C} \simeq 120$~K, $\rho_{xy}$ is described by the sum of a linear $R_{O} B$ term and a very small   $|4 \pi M R_{S } |$  $(\le 8 \times 10^{-6} ~\Omega$~cm below 100~K) which saturates at most 1~T.
  In this region, $R_{O}$ is obtained as the slope of $\rho_{xy}$ in the high-field region. 
  In region B, where $120 < T < 200$~K, $\rho_{xy}$ shows nonlinear behavior relative to the magnetic field.
   The initial slopes of all curves (for 140, 160, and 180~K) coincide with that for 200~K. However, the observed $\rho_{xy}$ curves cannot be fitted by equation~(1), suggesting that the nonlinear behavior of $\rho_{xy}$ does not come from the anomalous Hall effect. 
   We believe this behavior is caused by a change in the carrier density under the magnetic fields, as we explain later. 
   With negligibly small $R_{S }$, $R_{O}$ in magnetic field $B$ is calculated by $R_{O}=\partial \rho_{xy} /\partial B$ and the initial slope gives the "zero-field carrier density". 
   In region C, where $T \ge 200$~K, $\rho_{xy}$ is a linear function of $B$. 
   $R_{O}$ is calculated by $R_{O}=\partial \rho_{xy} /\partial B$, neglecitng $R_{S}$ term.

In Fig. 2, we summarize the $T$-dependence of the ordinary Hall coefficient $R_{O}$ and the free electron carrier density, $n=R_{O}/e$, under magnetic fields of 0 and 8~T.
In the lower part of Fig. 2, electrical resistivity, $\rho$, under magnetic fields of 0 and 8~T is also shown.
 In region A and C, $n$ is independent of the magnetic field.
In region B, $n$ apparently increases with the higher magnetic field. 
The zero field carrier density increased by a factor of about six from $n \sim 7 \times 10^{18}~$cm$^{-3}$ (0.001 electrons/f.u.) to $n \sim 5 \times 10^{19}~$cm$^{-3}$ (0.005 electrons/f.u.) across $T_{C}$ as the temperature was lowered. 
This suggests that the carrier density change significantly contributes to the conductivity change near $T_{C}$.
  Considering that conductivity increases by a factor of nearly 30 near $T_{C}$, the observed carrier increase would increase the conductivity by the same factor of six near $T_{C}$, and the remaining conductivity change should be attributed to a five-fold decrease in the carrier scattering, in a free electron model.
  Our preliminary optical conductivity data also support this carrier density change at $T_{C}$~\cite{ref17}.
   Around $T_{C}$ (in region B), $n$ appears to increase when a magnetic field is applied as indicated by the nonlinear field dependence of $\rho_{xy}$.
    The relationship between $n$ and $B$ is shown in Fig. 3.
     The zero-field carrier densities in this region are almost the same value as that observed at 200~K. 
     The carrier densities at 140 and 160~K increase to $2.3 \times 10^{19}$~cm$^{-3}$ and $1.2 \times 10^{19}$~cm$^{-3}$, respectively under 8 T.
 We thus concluded that negative CMR, i.e. a marked increase in conductivity caused by applying a magnetic field around $T_{C}$, is caused by increased $n$, as well as by the suppression of the ferromagnetic critical scattering. 
 
A possible mechanism of the carrier-density change in Tl$_{2}$Mn$_{2}$O$_{7}$ is a conduction-band-edge shift.
 As $T$ decreases, the ferromagnetic ordering occurs separately from the conduction process, due to the superexchange interaction of Mn localized moments (Mn$^{4+}$) {\it via} O and Tl ions.
  With the development of spontaneous magnetization, the internal magnetic field created by the Mn moments enhances the polarization of the conduction band through the strong $s$-$d$ interaction between conduction electrons and the localized Mn moments.
   As a result, the edge of the conduction band is lowered, and carrier density $n$ increases, significantly increasing conductivity.
 The carrier density change under a magnetic field is understood in the same manner, i.e. the applied magnetic field causes the conduction-band-edge shift.
   Under a magnetic field, in this model, $n$ is expected to increase following the relationship of $M^{3/2}$ versus $B$ within the range of the effective mass approximation near the bottom of a parabolic band.
    This is appropriate for Tl$_{2}$Mn$_{2}$O$_{7}$, which has a very small number of carriers with a nearly free electron mass in a parabolic conduction band~\cite{ref14a,ref14}.
 As shown in the inset of Fig. 3, the observed behavior of $n(B)$ is similar to the correlation between $M^{3/2}$ and $B$ at each temperature.
 Therefore, the conduction-band edge shifts downwards by changing either $T$ or $B$, giving rise to drastic changes in the transport properties. 
 This scenario is similar to that proposed for ferromagnetic semiconductors, europium calcogenides~\cite{ref18,ref19} or $n$-type CdCr$_{2}$Se$_{4}$~\cite{ref20}, and in sharp contrast with that for perovskite manganese oxides.
  In the perovskite crystals~\cite{ref21,ref22} and films~\cite{ref23,ref24}, several authors claimed that the critical magnetic scattering and the polaron formation, i.e. the mobility change, play important roles to increase in electrical resistivity above $T_{C}$ and CMR behaviors, altough the field-independent Hall mobility was observed in single-crystal La$_{2/3}$(Ca, Pb)$_{1/3}$MnO$_{3}$  within the CMR regime~\cite{ref22}. 
 In addition, obvious $R_{S }$, which has a sign opposite that of $R_{O}$, was observed in the ferromagnetic metallic state.

The $T$-dependence of thermopower, $S$, under magnetic fields of 0, 2, 5, and 7~T are shown in Fig. 4.
 The observed $S$ is negative throughout the measured $T$ and $B$ ranges. 
 This is consistent with Hall resistivity measurements, which revealed conduction with electron character. 
 Below $T_{C}$, $|S|$ is linearly dependent on $T$, which is typical behavior of normal metals. 
Around $T_{C} \simeq $~120~K, $|S|$ increases abruptly as temperature rises, and again above $T_{C}$ it shows a linear dependence on $T$ with a different slope.
Anomalies were observed around 280~K.
Since Hall resistivity and electrical resistivity~\cite{ref13} also show anomalies around this temperature, a change in the electronic structure may have occurred, probably due to spin frustration effects common in the pyrochlore structure.
The details of these anomalies are not yet clear. Since $dS/dT$ includes the $n^{-3/2}$ term in the simplest Mott's formula for metals, the change in the slope of $S$ indicates that $n$ increases below $T_{C}$. 
The relatively large observed values of $S$ compared with normal metals are consistent with the small carrier concentration in this material. 
In the case of La$_{1-x}$Sr$_{x}$MnO$_{3}$, on the other hand, the $T$- and $B$ -dependence of $S$ is complicated~\cite{ref25}.
 For a moderately doped crystal $(x=0.25)$, the sign reversal was observed around $T_{C}$ by changing either $T$ or $B$. 
 Even at temperatures far below $T_{C}$, $S$ did not show a liner dependence on $T$. 

The magneto-thermopower defined as $\Delta S=S(B)-S(0)$  is shown in the inset of Fig. 4.
Around $T_{C}$, a huge magneto-thermopower is observed, which is related to the carrier density change caused by an applied magnetic field.
 Above $T_{C}$, a substantially large magneto-thermopower remains.
  This is probably caused by the suppression of large thermal spin fluctuation by applied magnetic fields, because $n$ is independent of the magnetic field in this region.

We then analyzed the $T$-dependence of the thermopower. Since $S$ is strongly affected by the shape of the Fermi surface and the energy dependence of the relaxation time, it is difficult to write out its $T$- and $B$-dependence, in general.
 We believe, however, that the $T$-dependence of thermopower for Tl$_{2}$Mn$_{2}$O$_{7}$ can be described through an effective mass approximation, because Tl$_{2}$Mn$_{2}$O$_{7}$ has a simple nearly spherical Fermi surface~\cite{ref14a,ref14}. 
 In this model, the $T$-dependence of thermopower is written as,
 \begin{equation}
\displaystyle  S = - \frac{1}{eT} \left( \dfrac{\displaystyle \int \tau E^{\frac{5}{2}} \dfrac{\partial f_{0}}{\partial E}\,dE}{\displaystyle \int \tau E^{\frac{3}{2}} \dfrac{\partial f_{0}}{\partial E}\,dE}-E_{F} \right),
\end{equation}
 where $E_{F}$ is the Fermi energy, $\tau$ is the relaxation time, and $f_{0}$ is the Fermi-Dirac distribution function~\cite{ref26}. 
  We assume the relaxation time is written as
$\tau(E) = \tau_{0} E ^{\xi} $  
in the usual manner. 
For scattering by acoustic phonons, the value for $\xi$ is $-1/2$, and $\xi=3/2$ is applied for scattering by ionized impurities or optical phonons. 
With the expression of thermopower, we can calculate its $T$-dependence as shown in Fig. 5. 
In the calculation, $n$ changes with changing temperature as obtained from the Hall effect measurement. 
With the calculated results for $\xi=-1/2$, we can reproduce the observed behavior of $S$ below about 200~K, particularly, the abrupt increase in $|S|$ around $T_{C}$.
 It is, therefore, reasonable to consider that the observed change in $|S|$ is mainly caused by the change in the carrier concentration.
  Above 200~K, the experimental results deviate from the $\xi=-1/2$ curve and approach the $\xi = 3/2$ curve, suggesting a change in the scattering mechanism. 
  In this temperature region, the thermal excitation of optical phonons appears to become an important cause of scattering.

In conclusion, by investigating the magneto-transport phenomena of Tl$_{2}$Mn$_{2}$O$_{7}$, we found that the carrier dendity is increased by either lowering temperature or increasing the magnetic field near $T_{C}$. 
This explains the transport properties including CMR in Tl$_{2}$Mn$_{2}$O$_{7}$, and is in sharp contrast with perovskite manganese oxides.
We believe the carrier density is inceased by a conduction-band-edge shift, which is caused by the development of magnetization through the $s$-$d$ interaction between localized Mn moments and $s$-like conduction electrons.

\begin{figure}
\caption{
Magnetic-field dependence of the Hall resistivity at various temperatures.
}
\label{Fig1}
\end{figure}

\begin{figure}
\caption{
Temperature dependence of the ordinary Hall coefficient, $R_{O}$, and the free-electron carrier density, $n$. (top)
 Closed circles represent the zero field carrier density, open squares represent the carrier density at 8~T.  Lines are guides for the eye. 
The electrical resistivity under magnetic fields of 0 and 8~T is also shown. (bottom)
}
\label{Fig2}
\end{figure}

\begin{figure}
\caption{
Magnetic field dependence of free electron carrier density in region B.  In the inset, $M^{3/2}$ is plotted against $B$.
}
\label{Fig3}
\end{figure}

\begin{figure}
\caption{
Temperature dependence of thermopower measured under magnetic fields of  0, 2, 5 and 7~T. 
The inset shows magneto-thermopower defined by $\Delta S=S(B)-S(0)$. 
} 
\label{Fig4}
\end{figure}

\begin{figure}
\caption{
Temperature dependence of thermopower in the effective mass approximation. $S$ denoted by the open squares is calculated with $\xi=-1/2$, and $S$ for the closed squares is obtained with $\xi=3/2$. 
The closed circles are experimental data. }
\label{Fig5}
\end{figure}


\begin{references}
\bibitem[*]{byline}  To whom correspondence should be addressed.
\end{references}
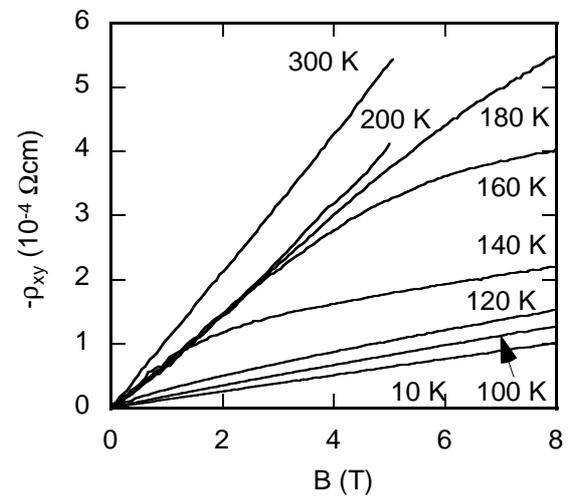
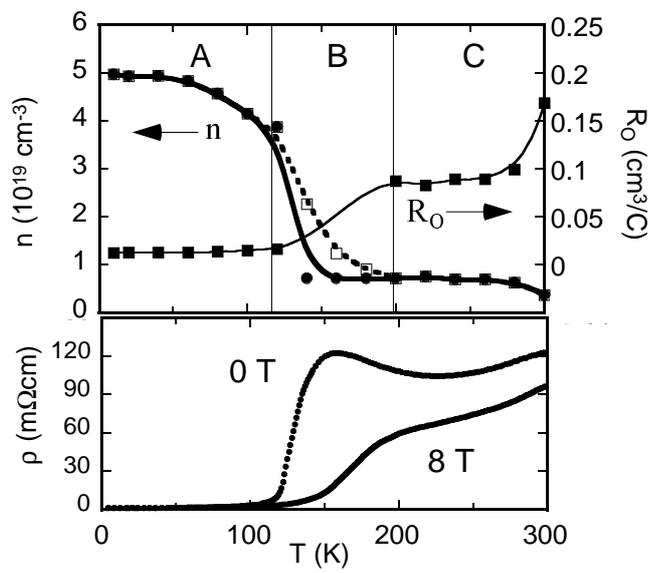
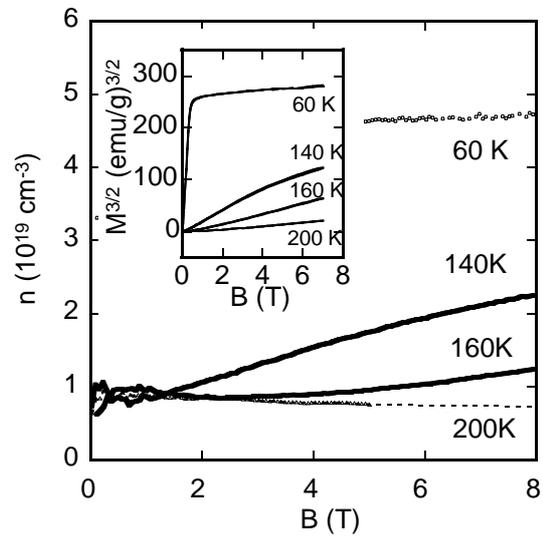
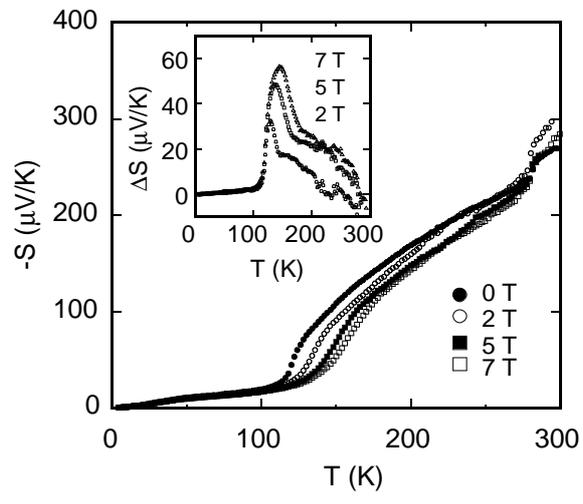
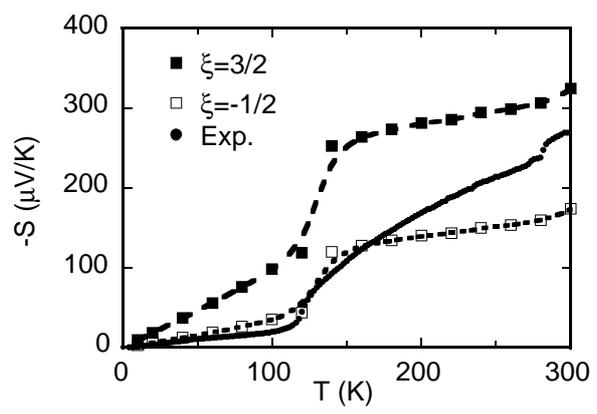
\end{document}